\documentclass[12pt]{article}
\pdfoutput=1

\usepackage{fixltx2e} 
\usepackage{cmap} 
\usepackage{ifthen}                                                                                                                           
\usepackage[T1]{fontenc}                                                                                                                      
\usepackage[utf8]{inputenc}                                                                                                                   
\usepackage{color}                                                                                                                            
                                                                                                                                              
\usepackage{mathptmx} 
\usepackage[scaled=.90]{helvet}                                                                                                               
\usepackage{courier}                                                                                                                          

\usepackage{graphics,amssymb,float}
\usepackage{xspace}
\usepackage{placeins}
\usepackage{rotating}
\usepackage{dcolumn}
\usepackage{bm}
\usepackage{cite}
\usepackage{amsmath}
\usepackage{indentfirst} 
\usepackage{tikz}
\usepackage{longtable,multirow}
\usetikzlibrary{shapes,arrows}

\def\eg{{\it e.g.}}

\newcommand{\sigmaXBF}{\mbox{\ensuremath{\sigma\times\mathcal{B}}}\xspace}

\def\lsim{\mathrel{\raise.3ex\hbox{$<$\kern-.75em\lower1ex\hbox{$\sim$}}}}
\def\gsim{\mathrel{\raise.3ex\hbox{$>$\kern-.75em\lower1ex\hbox{$\sim$}}}}
\def\ifmath#1{\relax\ifmmode #1\else $#1$\fi}

\def\Etmiss{$\not\!\!E_T$}

\def\smodels{{SModelS\,v1.0}}

\ifthenelse{\isundefined{\hypersetup}}{
  \usepackage[colorlinks=true,linkcolor=blue,urlcolor=blue]{hyperref}
  \urlstyle{same} 
}{}
\hypersetup{
  pdftitle={Installation and Deployment},
}

\begin{document}

\thispagestyle{empty}

\begin{center}

\begin{flushright}
LPSC14295\\
HEPHY-PUB 945/14
\end{flushright}

\vspace*{2cm}
{\LARGE\bf SModelS v1.0: a short user guide} 

\vspace*{1.6cm}

\renewcommand{\thefootnote}{\fnsymbol{footnote}}

{\large 
Sabine~Kraml$^{1}$,  
Suchita~Kulkarni$^{1,2}$,  
Ursula~Laa$^{1,2}$,  
Andre~Lessa$^{3}$,\\[1mm]  
Veronika~Magerl$^{2}$,  
Wolfgang~Magerl$^{2}$,  
Doris~Proschofsky-Spindler$^{2}$\footnote{Present address: Leopoldsgasse 21/2/11, 1020 Wien},\\[1mm] 
Michael~Traub$^{2}$,  
Wolfgang~Waltenberger$^{2}$  
} 

\renewcommand{\thefootnote}{\arabic{footnote}}

\vspace*{1cm} 

{\normalsize \it 
$^1\,$Laboratoire de Physique Subatomique et de Cosmologie, 
Universit\'e Grenoble-Alpes, CNRS/IN2P3, 53 Avenue des Martyrs, F-38026 Grenoble, France\\[2mm]
$^2\,$Institut f\"ur Hochenergiephysik,  \"Osterreichische Akademie der Wissenschaften,\\ Nikolsdorfer Gasse 18, 1050 Wien, Austria\\[2mm]
$^3\,$Instituto de F\'isica, Universidade de S\~ao Paulo, S\~ao Paulo - SP, Brazil
}

\vspace*{1cm}

Email: smodels-users@lists.oeaw.ac.at

\vspace*{1cm}

\begin{abstract}
SModelS is a tool for the automatic interpretation of simplified-model results from the LHC. 
Version~1.0 of the code is now publicly available. 
This document provides a quick user guide for installing and running SModelS\,v1.0.
\end{abstract}

\end{center}

\setcounter{page}{0}\clearpage 

\section{Introduction}\label{sec:intro}

\begin{figure}[h!]\centering
\includegraphics[width=0.9\textwidth]{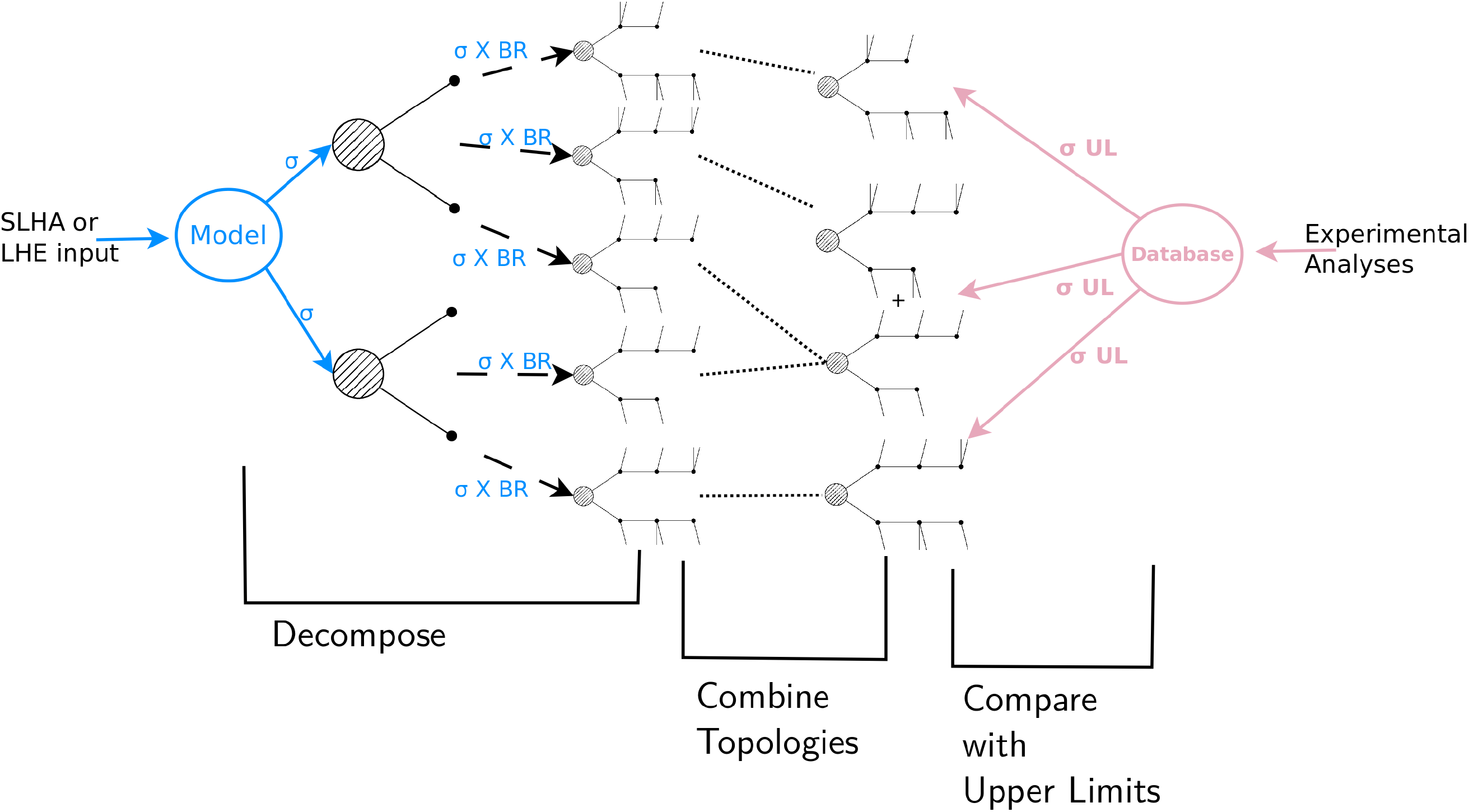}
\caption{\label{fig:scheme}
Schematic view of the working principle of {\tt SModelS}.}
\end{figure}

SModelS~\cite{Kraml:2013mwa} is an automatised tool for interpreting simplified-model results from the LHC.
It is based on a general procedure to decompose Beyond the Standard Model (BSM) collider signatures presenting a 
$Z_2$ symmetry into Simplified Model Spectrum (SMS) topologies.
Our concrete implementation currently focusses on supersymmetry (SUSY) searches with missing transverse energy, 
for which a large variety of SMS results from ATLAS and CMS are available.
The main ingredients of SModelS are
\begin{itemize}
\item the decomposition of the BSM spectrum into SMS topologies,
\item the database of experimental SMS results,
\item the interface between decomposition and the results database. 
\end{itemize}
The working principle is illustrated schematically in Fig.~\ref{fig:scheme}.

Version 1.0 of SModelS is now publicly available. This document provides a short user guide for the installation and for running \smodels. More details can be found in the primary physics publication \cite{Kraml:2013mwa} and in the documentation provided with the code (an html manual, which is also available online at \cite{smodels:wiki}). 
In case of problems using \smodels, we kindly ask the user to contact the authors at smodels-users@lists.oeaw.ac.at.

\clearpage
\section{Download and Installation}

SModelS is a Python library that requires Python version 2.6 or later (but not version 3). 
Internally, \smodels\ uses Pythia\,6.4.27~\cite{Sjostrand:2006za}, NLL-fast versions 1.2 and 2.1 \cite{nllfast} 
(see also \cite{Beenakker:1996ch,Beenakker:1997ut,Kulesza:2008jb,Kulesza:2009kq,Beenakker:2009ha,Beenakker:2010nq,Beenakker:2011fu}) 
and a modified version of PySLHA~\cite{Buckley:2013jua}. 
It has been tested on Ubuntu 14.X, Scientific Linux (CERN) 5 and 6, 
as well as on Mac Os X 10.9 and 10.10. 

The \smodels\ package can be downloaded from \cite{smodels:wiki}. Unpacking the tar ball with 
\begin{quote}
  {\tt tar -zxvf smodels-v1.0.tgz}
\end{quote}
creates the directory {\tt smodels-v1.0}, where the code (subdirectory {\tt smodels}) and the results database (subdirectory {\tt smodels-database}) are located.

For installation, SModelS makes use of Python's setuptools. 
On most machines (apart from SL and SLC)
\begin{quote}
  {\tt python setup.py install}
\end{quote}
inside the {\tt smodels-v1.0} directory should install the entire project, resolving automatically the external Python 
dependencies and compiling the internal Pythia\,6 and NLL-fast versions using {\tt gfortran}. 
Note that depending on the path specified in {\tt setup.py} you may need to do the installation as superuser. 

More details on the installation procedure, external dependencies and in particular instructions for installation 
on Scientific Linux can be found in the README file in {\tt smodels-v1.0}. 

\section{Citation}

When you use results obtained with \smodels\ in a publication, please cite this document as well as 
the original SModelS paper~\cite{Kraml:2013mwa}, Pythia\,6.4~\cite{Sjostrand:2006za}, 
NLL-fast \cite{Beenakker:1996ch,Beenakker:1997ut,Kulesza:2008jb,Kulesza:2009kq,Beenakker:2009ha,Beenakker:2010nq,Beenakker:2011fu}, and PySLHA~\cite{Buckley:2013jua}. 
For convenience, these citations are provided in bibtex format in the \smodels\ distribution.

\section{Using SModelS}

\subsection{Basic input}\label{sec:input}

\smodels\ can be used with two forms of inputs

\begin{itemize} 
\item SLHA (SUSY Les Houches Accord)~\cite{Skands:2003cj} files containing masses,\\ 
branching ratios and cross-sections, or 
\item LHE (Les Houches Event)~\cite{Alwall:2006yp} files containing parton-level events.
\end{itemize} 

The SLHA format is usually more compact and best suited for supersymmetric models. 
On the other hand, a LHE file can always be generated for any BSM model 
(through the use of your favorite Monte Carlo generator).\footnote{\smodels\ can easily be used for non-SUSY models as
long as they present a $Z_2$-type symmetry. However, it is the responsibility of the user to make sure that the SMS results in the database actually apply to the model under consideration. In this context, see also the caveats in \cite{Kraml:2013mwa}.} In this case, however,
the precision of the results is limited by the MC statistics used to generate the file.

In the case of {\bf SLHA input}, the production cross sections for the BSM states also have 
to be included as SLHA blocks, according to the SLHA cross-section format \cite{SLHA-xsections}. 
For the MSSM and some of its extensions, the cross sections can be conveniently calculated and added to the SLHA input file by means of the internal SModelS' {\tt xseccomputer} described in the tools section of the html manual. 

In the case of {\bf LHE input}, the total production cross-section as well as 
the center-of-mass energy should be listed in the <init></init> block, according to the standard LHE format.
Moreover, all the $Z_2$-even BSM particles (\eg, additional Higgs states) should be set as stable, since
in \smodels\ they are effectively considered as final states.

Besides information about the masses and branching ratios ($\mathcal{B}$), the user must also {\bf define} which of the particles are 
{\bf $Z_2$-odd states} and which are $Z_2$-even. This is done in the {\tt particles.py} file, where some default values (for SM and MSSM particles) are already loaded. Finally, if the user wants to check the SLHA input file for possible issues using SModelS' {\tt slhachecker}, it is also necessary to define the BSM particle quantum numbers in {\tt particles.py}.

\subsection{runSModelS.py}

For the first-time user, SModelS ships with a command-line tool, {\tt runSModelS.py}, 
which  covers several different applications of the SModelS functionalities. 
These functionalities include detailed checks of input SLHA or LHE files,  
running the SMS decomposition, evaluating the theory predictions for the input model 
and comparing them to the experimental limits available in the database, 
determining the most important missing topologies and printing a summary text file.

The usage is
\begin{quote}
  {\tt runSModelS.py -f INPUTFILE [-p PARAMETERFILE] [-o OUTPUTFILE]}
\end{quote}

\noindent Some comments are in order. 
\begin{itemize}
\item The INPUTFILE can be an SLHA or LHE file as explained in Section~\ref{sec:input}.
\item The PARAMETERFILE controls the basic options and parameters used by {\tt runSModelS.py}. 
An example including all available parameters together
with a short description, is provided as {\tt parameters.ini}. 
If no parameter file is specified, the default parameters stored in {\tt etc/parameters\_default.ini} are used.
Note that the input type (SLHA or LHE) needs to be properly specified in the parameter file.
\item If no OUTPUTFILE is specified, the file output will be printed to {\tt summary.txt}.
\end{itemize}

\clearpage
\subsection{Default output}

The results of {\tt runSModelS.py} are printed to the screen and to the output (summary) file. 
The level of detail is controlled via the parameters file. 
The screen output comprises
\begin{itemize}
\item a full list of the topologies generated by the decomposition procedure\\ (if {\tt printDecomp = True}),
\item a list of all the analyses considered (if {\tt printAnalyses = True}),
\item a list of all the theory predictions obtained and the corresponding upper limits from the experiments (if {\tt printResults = True}),
\item possible warnings or error messages. 
\end{itemize}

The file output contains status flags for the input file and the decomposition, indicating possible problems.
The status flags are followed by the name of the input file, basic information on the run parameters, and the version of the database used.  This looks as follows:
{\footnotesize
\begin{verbatim}
Input status: 1
Decomposition output status: 1 #decomposition was successful
#Input File: inputFiles/slha/gluino_squarks.slha
#maxcond = 0.2
#minmassgap = 5.
#sigmacut = 0.03
#Database version: GrenobleNov2014 (27/11/2014)
\end{verbatim} }
The main part of the output file is then the list of analyses which constrain the input model. 
For each analysis, its ID, Tx name,\footnote{The {\bf Tx names} are explained in the SMS dictionary on \href{http://smodels.hephy.at/wiki/SmsDictionary}{http://smodels.hephy.at/wiki/SmsDictionary}.} centre-of-mass energy and the amount of condition violation are given, 
followed by the predicted signal cross section and the 95\% CL experimental upper limit on it. 
The last entry in the line is the ratio $r$ of the signal cross-section and the upper limit, 
$r=\sigma(\textrm{predicted})/\sigma(\textrm{excluded})$, where $\sigma$ effectively means 
$\sigmaXBF$ or the weight of the topology. 
A value of $r\ge 1$ means that the input model is likely excluded by the corresponding analysis.
Concretely, this looks like this:  
{\footnotesize
\begin{verbatim}
#Analysis  Tx_Name  Sqrts  Cond. Violation  Theory_Value(fb)  Exp_limit(fb)  r

 CMS-PAS-SUS-13-019         T2  8.00E+00  0.0  1.773E+00  3.762E+00  4.714E-01
#[[['jet']],[['jet']]]
--------------------------------------------------------------------------------
     CMS-SUS-13-012         T2  8.00E+00  0.0  1.773E+00  6.099E+00  2.907E-01
#[[['jet']],[['jet']]]
--------------------------------------------------------------------------------
 ATLAS-SUSY-2013-12     TChiWZ  8.00E+00  0.0  1.847E+01  3.301E+02  5.595E-02
#[[['W']],[['Z']]]
--------------------------------------------------------------------------------
\end{verbatim} }

\noindent
In this example, the topologies tested by the analysis are shown in bracket notation just below the analysis entry, 
here \eg\ {\tt [[['jet']],[['jet']]]}. This can be turned off by setting {\tt addConstraintInfo = False} in the parameter file.
The last line of this block reports the maximum value of $r$:
{\footnotesize
\begin{verbatim}
The highest r value is 4.71E-01 
\end{verbatim} }

\noindent
The detailed breakdown can be switched off by setting 
{\tt expandedSummary=False}; in this case only the most constraining analysis (the one with the maximum $r$ value) is printed.

Finally, if {\tt findMissingTopos = True},  a list of the missing topologies (in bracket notation) and their cross sections at the given $\sqrt{s}$ is also included. This list is ordered from high to low cross sections; per default only the 10 leading ones are printed.
An example is shown below:
{\footnotesize
\begin{verbatim}
Missing topologies with the highest cross-sections (up to 10):
Sqrts (TeV)   Weight (fb)        Element description
8.00E+00   1.567E+01    #                 [[[jet],[W]],[[jet,jet],[W]]]
8.00E+00   1.395E+01    #       [[[jet],[jet,jet],[W]],[[jet,jet],[W]]]
\end{verbatim} }

\noindent
For more details, including an explanation of the bracket notation, we refer the user to the html manual. 

\subsection{Using \smodels\ as a Python library}

Although {\tt runSModelS.py} provides the main SModelS features with a command line interface, 
users familiar with Python and the SModelS language may prefer to write their own main program, 
using \smodels\ as a Python library. 
A simple example code for this purpose is provided as {\tt Example.py} in the \smodels\ distribution. 
Further examples are presented in the ``More Examples'' section of the manual.

\section{List of experimental results included in the database}

The \smodels\ database comprises a large number of SMS results from ATLAS and CMS 
SUSY searches at $\sqrt{s}=8$~TeV. Most of them are for full luminosity. 
The 7~TeV results will be included in a later release together with more 8 TeV results and we
plan to update the database in a regular basis in the future.
Concretely, in the default run mode, the following results are considered:\footnote{The luminosity 
L is given in  [fb$^{-1}$]; the shorthand notation T1tttt(off) stands for ``T1tttt and T1ttttoff'' (on-shell and off-shell, respectively).
Likewise  TChiWZ(off) denotes ``TChiWZ and TChiWZoff''.}
 
\bigskip

\noindent
\begin{tabular}{l l c l l}
\bf ID & \bf short description & \bf L & \bf Ref. & \bf Tx names \\
\hline
ATLAS-SUSY-2013-02 & 0 leptons + 2--6 jets + \Etmiss & 20.3 & \cite{Aad:2014wea} &  T1, T2\\

ATLAS-SUSY-2013-04 & 0 leptons + {\small $\ge$}7--10 jets + \Etmiss & 20.3 & \cite{Aad:2013wta} & T1tttt  \\

ATLAS-SUSY-2013-05 & 0 leptons + 2 b-jets + \Etmiss & 20.1 & \cite{Aad:2013ija} & T2bb \\

ATLAS-SUSY-2013-11 & 2 leptons ($e,\mu$) + \Etmiss & 20.3 & \cite{Aad:2014vma} &  TChiWZ, TSlepSlep\\

ATLAS-SUSY-2013-12 & 3 leptons ($e,\mu,\tau$) + \Etmiss & 20.3 & \cite{Aad:2014nua} & TChiWH, TChiWZ(off)\\

ATLAS-SUSY-2013-14 & 2 taus + \Etmiss & 20.3 & \cite{Aad:2014yka} &  TStauStau\\

ATLAS-SUSY-2013-15 & 1 lepton + 4(1 b-)jets + \Etmiss & 20.3 & \cite{Aad:2014kra} &  T2tt, T2bbWW\\

ATLAS-SUSY-2013-19 & 2 leptons + (b)jets + \Etmiss & 20.3 & \cite{Aad:2014qaa} &  T2tt, T2bbWW, \\
&&&&T6bbWW\\
\hline

ATLAS-CONF-2012-105 & 2 SS leptons + {\small $\ge$}4 jets + \Etmiss & 5.7 & \cite{ATLAS-CONF-2012-105} & T1tttt \\

ATLAS-CONF-2013-007 & 2 SS leptons + 0--3 b-jets + \Etmiss & 20.7 & \cite{ATLAS-CONF-2013-007}  & T1tttt\\

ATLAS-CONF-2013-024 & 0 lepton + 6 (2 b-)jets + \Etmiss & 20.5 & \cite{ATLAS-CONF-2013-024}    & T2tt\\

ATLAS-CONF-2013-061 & 0--1 leptons + {\small $\ge$}3 b-jets + \Etmiss & 20.1 &  \cite{ATLAS-CONF-2013-061} &  T1bbbb, T1tttt\\

ATLAS-CONF-2013-065 & 2 leptons + (b)jets + \Etmiss & 20.3 &  \cite{ATLAS-CONF-2013-065} &  T2tt\\
\hline
\end{tabular}{}

\noindent
\begin{tabular}{llcll}
\bf ID & \bf short description & \bf L & \bf Ref. & \bf Tx names \\
\hline
CMS-SUS-12-024 & 0 leptons + {\small $\ge$}3(1 b-)jets + \Etmiss & 19.4 &  \cite{Chatrchyan:2013wxa}&  T1bbbb, T1tttt(off), T5tttt\\

CMS-SUS-12-028 & jets + \Etmiss, $\alpha_T$& 11.7 &  \cite{Chatrchyan:2013lya}&  T1, T1bbbb, T1tttt, T2, T2bb\\

CMS-SUS-13-002 & {\small $\ge$}3 leptons (+jets) + \Etmiss & 19.5 &  \cite{Chatrchyan:2014aea}&  T1tttt\\

CMS-SUS-13-006 & EW productions with  & 19.5 &  \cite{Khachatryan:2014qwa} &  TChiWZ(off), TSlepSlep, \\
& decays to leptons, W, Z,  &&& TChiChipmSlepL,\\
& and Higgs &&& TChiChipmSlepStau \\
CMS-SUS-13-007 & 1 lepton + {\small $\ge$}2 b-jets + \Etmiss & 19.3 &  \cite{Chatrchyan:2013iqa}&  T1tttt(off)\\

CMS-SUS-13-011 & 1 lepton + {\small $\ge$}4(1 b-)jets + \Etmiss & 19.5 &  \cite{Chatrchyan:2013xna}& T2tt, T6bbWW \\

CMS-SUS-13-012 & jet multiplicity + $\not\!\!H_T$ & 19.5 &  \cite{Chatrchyan:2014lfa}&  T1, T1tttt(off), T2\\

CMS-SUS-13-013 & 2 SS leptons + (b-)jets + \Etmiss & 19.5 &  \cite{Chatrchyan:2013fea}&  T1tttt(off),\\

\hline

CMS-PAS-SUS-13-008 & 3 leptons + (b)jets + \Etmiss & 19.5 &  \cite{CMS-PAS-SUS-13-008}&  T6ttWW, T1tttt\\

CMS-PAS-SUS-13-016 &  2 OS leptons\,+ {\small $\ge$}4(2b-)jets + \Etmiss & 19.7 &  \cite{CMS-PAS-SUS-13-016}&  T1tttt(off)\\

CMS-PAS-SUS-13-018 & 1--2 b-jets + \Etmiss, $M_{CT}$ & 19.4 &  \cite{CMS-PAS-SUS-13-018}&  T2bb\\

CMS-PAS-SUS-13-019 & hadronic $M_{\rm T2}$ & 19.5 &  \cite{CMS-PAS-SUS-13-019}&  T1, T1bbbb, T1tttt(off),\\
&&&& T2, T2tt, T2bb\\

CMS-PAS-SUS-14-011 & razor with b-jets & 19.3  &  \cite{CMS-PAS-SUS-14-011}&  T1bbbb, T1tttt(off), T2tt \\
\hline
\end{tabular}

\vspace*{8mm}

\noindent 
A bibtex file, database.bib, is available in the smodels-database/8TeV folder, which can conveniently be used for citing these analyses.

The database contains moreover the entries of several preliminary results (ATLAS conference notes or CMS analysis summaries) 
which were superseded by a publication. These results are not listed here, and are not used 
when running {\tt runSModelS.py}. They can be activated by 
\begin{quote}
{\tt smsAnalysisFactory.load(useSuperseded=True)} 
\end{quote}
when using \smodels\ as a Python library. 
(For more details, see "How to load the database" in the ``More Examples'' section of the html manual.)
Such superseded results can be useful, \eg, for comparing with older studies. 
It is, however, strongly discouraged to use them in any other case. 

Last but not least we note that for topologies with more than one step in the decay chain, \eg\ charginos decaying through  intermediate sleptons, or stops decaying into bottom plus chargino followed by the chargino decay into the lightest neutralino, we need several (more than one) mass planes in order to interpolate between them. 
{\em Whenever only one mass plane is provided, the result is not useful for our purpose and thus not included in the database.} 
 
\section{Conclusions}

We presented the first public release of SModelS, an automatic tool for interpreting 
simplified-model results from the LHC in generic models possessing a $Z_2$ symmetry. 
\smodels\ consists of a SMS decomposition procedure, a database of SMS cross section upper limits from ATLAS and CMS, and an interface between these two components to confront the theoretical predictions of BSM models with the experimental results. The database of v1.0 comprises results from 13 ATLAS and 13 CMS SUSY searches at 8~TeV, 
corresponding to 21 ATLAS and 41 CMS results when counting the individual results (Tx names) in the different publications.   
In this document, we provided the basic instructions for installing and running \smodels\ and understanding its output.  
More detailed explanations are given in the html manual which comes as part of the \smodels\ distribution. 

We hope that \smodels\ will be a useful tool for the High-Energy Physics community and contribute to the legacy of the LHC results. Here note that \smodels\ can be used not only to test whether a particular scenario is excluded by the recent LHC  results --- it can also be conveniently used to classify untested regions, missing topologies and new signatures that might be interesting to look for. 

This said, the release of \smodels\ is but the beginning of the story. 
Several extensions and improvements are already in development. 
One such extension will be a more extensive database of experimental results, ideally comprising also non-SUSY searches. This, however, depends also on the co-operation of the ATLAS and CMS search groups in making the relevant information available. Another extension will be the inclusion of efficiency maps. This will be a major upgrade of the code. 
Further plans include ways to determine the most sensitive analysis and ways to combine results from different analyses, extensions to asymmetric branches (if suitable experimental results become available for this) and signatures without missing energy.

\section*{Acknowledgements} 

We thank the ATLAS and CMS SUSY groups for helpful discussions on their results, 
and in particular for providing (most of) the SMS cross section upper limits used here 
in digital format. 

This work is supported in part by the French ANR project {\sc DMAstroLHC}.  
Su.K.\ is supported by the ``New Frontiers'' program of the Austrian Academy of Sciences.
U.L.\ is supported by  the ``Investissements d'avenir, Labex ENIGMASS''. 
A.L.\ is supported by FAPESP; he acknowledges moreover the hospitality of LPSC Grenoble and of HEPHY Vienna.
V.M.\ is grateful for financial support by the FEMtech initiative of the BMVIT of Austria.


\bibliography{references}

\end{document}